\documentclass{iaus}
\usepackage{graphicx}

\title[IAU 275~~ Relativistic Jets at High Energies] 
{Relativistic jets at high energies}

\author[Amir Levinson]   
{Amir Levinson$^1$}

\affiliation{$^1$School of Physics and Astronomy, Tel Aviv University\\ Tel Aviv 69978, Israel
\\ email: {\tt levinson@wise.tau.ac.il} \\[\affilskip]}

\pubyear{2010}
\volume{275}  
\pagerange{119--126}
\jname{Jets at all Scales}
\editors{G. E.~Romero, R.~A.~ Sunyaev \& T.~Beloni, eds.}
\begin{document}

\maketitle

\begin{abstract}
A recent progress in the study of $\gamma$-ray jets is reviewed, with a focus on some theoretical interpretations of
the VHE emission from M87, and possibly other misaligned blazars; the connection between the GeV breaks exhibited 
by bright LAT blazars and opacity sources in the broad line region;
the consequences of the detection of GeV emission from GRBs to models of magnetic outflows; and the implications
of the thermal emission observed is some GRBs to dissipation of the outflow bulk energy. 
\keywords{Jets, blazars, gamma ray bursts.}
\end{abstract}

\firstsection 
\section{Introduction}
Observations by Fermi and the various TeV experiments, and advances in
numerical techniques have led to a progress in our understanding of relativistic jets:  i)  In M87,  combined VLBA and TeV 
data (Acciari et al. 2009) seem to indicate that the TeV emission is produced on horizon scales by either some magnetospheric process, 
or at the base of the VLBA jet.  The observational constraints raise interesting questions about the structure 
of the BH magnetosphere and the jet formation mechanism, that appear to be relevant also to other TeV AGNs.  ii) The LAT 
spectrum of blazars provide now a better than before probe of opacity sources on sub-parsec scales.  The data
reveal relationship between source power and spectral features that can be interpreted in terms of BLR properties.  
iii) Combined Fermi and TeV observations of VHE blazars may be used to probe intergalactic magnetic fields 
(e.g., Neronov et al. 2010; Tavecchio et al. 2010).  Attempts to 
derive constrains on the IGMF have been published, and although inconclusive yet, the results demonstrate the
potential in pursuing further these efforts.  iv)  Fermi LAT detections of several GRBs  indicate very high Lorentz factors in 
outflows having opening angles larger than the causality scale.  Those detections triggered  recent theoretical studies and 
numerical simulations of magnetic outflows.  The conclusion emerging from these studies is that processes beyond ideal MHD are crucial.
Hydrodynamical outflows, perhaps driven by neutrino annihilation, may provide an alternative. The nature of the GeV emission in GRBs is
yet an open issue.  v)  In several bursts a prominent, quasi-thermal spectral component has been detected, challenging the 
"standard model". It opens up the issue of dissipation and emission during the prompt phase.  In particular, part of the emission 
is likely produced behind relativistic radiation mediated shocks.  

In what follows, these issues are discussed in some greater detail.

\section{GeV-TeV emission from AGNs}
Over 700 AGNs, most of which are blazars, are listed in the first Fermi LAT catalogue (Abdo, et al. 2010c).   Around 3 dozens AGNs have 
been detected by various TeV experiments at energies above 100 GeV, with a spectrum extending up to 10 TeV in some cases. 
At least 50 percents of the TeV sources have GeV counterparts.   The LAT spectrum of many bright Fermi blazars exhibits curvature
or breaks at GeV energies (Abdo et al. 2010b), conceivably associated with attenuation by radiation emitted from highly ionized clouds in the broad line region (BLR).  

Eleven non-blazar AGNs, - 7 FRI radio galaxies and
 4 FRII radio sources - exhibit VHE ($>100$ MeV) emission (Abdo, et al. 2010a), and some of them, e.g., M87  (Wagner et al. 2009), 
 Cen A (Aharonian, et al. 2009), are also TeV sources.   
 According to the unified model, those radio galaxies are misaligned blazars, and the question then arises if the VHE photons
 seen represent side emission from the jet, or have a different origin.   In the former case, the high luminosities 
 measured in some of the objects may imply surprisingly small Doppler factors, or extremely high jet power.
 
It is conceivable that the misaligned blazars exhibit spectral components that are difficult to detect in blazars because they are overwhelmed 
by the beamed emission from jet.   Emission from a starved magnetosphere is one possibility (Levinson 2000).  Recent VLBA and TeV observations of 
M87 (Acciari et al. 2009) strongly motivates reconsideration of this scenario.   Below we propose that the variable TeV emission observed in M87 could be 
a manifestation of the jet formation process.

\subsection{Magnetospheric TeV emission and jet formation in M87}
Combined VLBA and TeV observations of M87 reveal a rapidly varying TeV emission that appears to be associated with 
the m.a.s VLBA jet.  The rapid flaring activity of the TeV source, with timescales $t=1t_{day}$ day as 
low as 1-2 days, implies a source size of $d\sim 4r_s t_{day}$ for a black hole mass
$M_{BH}=4\times10^9$ solar masses.  This, and the fact that the TeV emission appears to be correlated with the VLBA jet but not
with emission from larger scales (and in particular HST-1), motivates the consideration that the observed TeV rays originate from the
black hole magnetosphere.  A plausible magnetospheric process discussed in the literature 
is curvature and/or IC emission by particles, either hadrons or leptons, accelerating in 
a vacuum gap of a starved magnetosphere.   An alternative explanation is emission  from small regions located at 
larger radii, $r\sim 100r_g$, as, e.g., in the misaligned minijets model of  Gainnios et al. (2010).

\begin{figure}
\centering
\includegraphics[width=100mm,height=80mm]{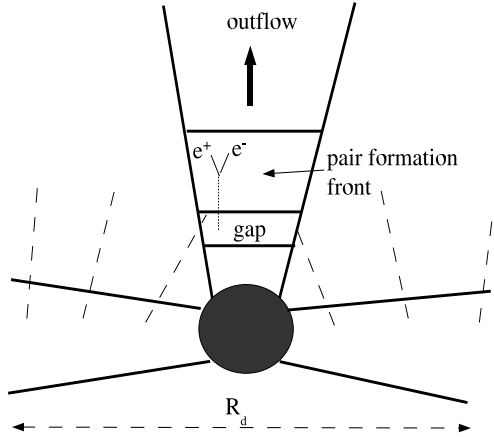}
\caption{\label{fig:gap}Schematic representation of the magnetosphere structure: A vacuum gap of height $h<r_g$ accelerates particles (electrons or positrons) to high Lorentz factors.  The gap is exposed to soft radiation emitted from a source of size $R_d>>r_g$.  Curvature emission and inverse Compton scattering of ambient radiation produce VHE photons with a spectrum extending up to $10^4$ TeV.   Photons having energies below a few TeV can escape freely to infinity.   Interactions of IC photons having energies well above 10 TeV with the ambient radiation initiate pair cascades just above the gap,  leading to a large multiplicity.  A force free outflow is established just above the pair formation front, and appears as the VLBA jet.   
Intermittencies of the cascade process give rise to the observed variability of the TeV emission, and the resultant force-free outflow, as indicated
by the morphological changes of the VLBA jet.}
\end{figure}

The presence of the VLBA jet implies that a force-free (or ideal MHD) flow is established on scales $< 100 r_g$, so that the magnetosphere 
is anticipated to be screened in the sense that the invariant ${\bf E} \cdot {\bf B}$ nearly vanishes everywhere.   However,
as in pulsar theory, there must be a plasma source that compensates for the loss of particles which escape
the system (both, to infinity and across the horizon) along the open magnetic field lines in the polar region.   The nature of this plasma source is poorly understood at present.   

The injection of charges into the magnetosphere may be associated with the accretion process.  Direct feeding
seems unlikely, as charged particles would have to cross 
magnetic field lines on timescale shorter than the accretion time in order to reach the polar outflow.  Free neutrons, if exist,
decay over length scale much smaller than the gravitational radius of the black hole. Annihilation of MeV photons produced in the radiative inefficient flow, may provide the required charge density, depending on accretion rate and other details. 

A possible plasma source is cascade formation in starved magnetospheric regions.  The size of the gap 
then depends on the conditions in the magnetosphere and the pair production opacity.   As shown elsewhere (Levinson 2000), 
vacuum breakdown by back-reaction is unlikely, as it requires magnetic field strength in excess of a few times $10^5$ G, higher than the equipartition 
value for Eddington accretion.  Pair production via absorption of TeV photons by the ambient radiation field is more likely.  However, 
the spectrum  of the VHE photons observed by HESS extends up to 10 TeV, and the assumption that these photons originate from the
magnetosphere (or even the VLBA jet) implies that the pair production opacity at these energies must not exceed unity.  This raises the question 
whether pair cascades in the magnetosphere can at all account for the multiplicity required to establish a force-free flow.

In an upcoming paper (Levinson \& Rieger, 2010, in preparation)  it is shown that inverse Compton scattering (IC)
of ambient photons by electrons accelerated in the gap can lead to
a large multiplicity, in excess of $10^3$, while still allowing photons at energies of up to a few TeV to freely escape the system.   The 
seed electrons are generated by the MeV emission from the RIAF. This requires
that the ambient radiation source contributing the opacity will have a characteristic size $R_d\sim 10^2 r_g$.  Preliminary results 
indicate that the electromagnetic cascade is 
initiated by IC photons having much higher energies, $\sim 10^3$ TeV, for which the $\gamma\gamma$ optical depth is much larger. 
It is found that the gap width is not much smaller than $0.1 r_g$.  The luminosity of the VHE photons produced in the gap,
roughly a fraction $(h/r_g)^2$ of the maximum BZ power, can easily account for the TeV luminosity observed by HESS and Fermi.  
Any intermittencies of the cascade 
formation process would naturally lead to variability of both, the magnetospheric TeV emission and the resultant force-free flow, as 
observed.  A schematic illustration of the model is presented in figure \ref{fig:gap}. 

This model may be applicable also to some of the other misaligned VHE blazars. 

\subsection{GeV breaks and opacity sources in blazars}
The nature of opacity sources in blazars is an issue of considerable interest, particularly in regards to the location of the VHE emission zones.
Much efforts have been devoted over the years to identify different opacity sources, and detailed calculations  of $\gamma\gamma$ attenuation 
by these sources are presented in the literature.  Recent Fermi observations have led to reconsideration of the effect of the broad line region on
the GeV spectrum of blazars.

Strong departure from a single power law appears to be a common feature in FSRQs (Abdo et al. 2010b).  The break energies, typically 
lying in the energy range 2-10 GeV, are too low to be due to absorption by Ly$\alpha$ radiation.   Recently, it has been pointed out (Putanen \& Stern 2010) that the LAT spectrum of FSRQs can be well reproduced by a double absorber model.  Detailed numerical calculations of 
the BLR spectrum for different ionization parameters indicates that the observed GeV breaks can be accounted for 
by $\gamma$-ray absorption through pair production on He II and H I recombination continua.  The exact location of the break energy depends on the ionization parameter, as described in Putanen \& Stern (2010).  The fact that GeV breaks are seen mainly in the brightest blazars 
is naturally explaned by the scaling of the BLR size with luminosity, $R_{BLR}\propto L^{1/2}$, obtained from reverberation 
mapping techniques (Kaspi et al. 2007), that implies $\tau_{\gamma\gamma}\propto L^{1/2}$.

\section{VHE emission and dissipation in GRBs}
Over a dozen GRBs, both long and short, have been detected thus far by  LAT onboard Fermi.  The origin of the GeV emission is
yet an open issue.  The flux level of the LAT emission and its long lasting light curve, suggest that this component is produced 
during the afterglow phase, although other explanations have been offered.   For sources that show sub-second 
variations of the GeV flux, e.g., 080916C, opacity arguments yield  high Lorentz factors, $\Gamma\sim 10^3$.  While various factors 
render those estimates uncertain (e.g., Yuan-Chuan et al. 2010),  it seems difficult to avoid the conclusion that the outflow reaches high Lorentz factors. 
Such high Lorentz factors posses a great challenge to jet models, as discussed in the following.

Another interesting result is that in some GeV bursts, e.g., GRB 090902B (Peer et al. 2010), a prominent quasi-thermal spectral component has been 
detected, whereas in others, e.g., 080916C, a fit by a Band spectrum is claimed to be satisfactory, and there appears to be no indication for thermal emission. 
This, presumably, reflects the conditions in the dissipation region. Below it is argued that in bursts that exhibit a thermal component a substantial 
fraction of the outflow bulk energy dissipates below the photosphere, in relativistic radiation mediated shocks, whereas in other bursts
most of the energy dissipates above the photosphere, behind collisionless shocks.  

\subsection{Formation and acceleration of magnetic jets}

Extraction of rotational energy from a rapidly spinning black hole by magnetic fields is a widely discussed jet production mechanism
 (e.g., Levinson \& Eichler 1993; Lyutikov \& Blandford 2003; Giannios \& Spruit 2005).  
Up to 30 percents of the black hole mass can be extracted from a maximally rotating hole, which is more than sufficient to account
for the energetics of essentially all  GRBs.  The rate at which the energy is extracted depends on the strength of the magnetic field 
in the vicinity of the horizon.  Recent numerical simulations confirm that a Blandford-Znajek process does operate in Kerr spacetime, and that for
reasonable assumptions about the progenitor star, sufficiently large power can be extracted in the form of an outflow.  
Moreover, a very low baryonic content is anticipated along horizon threading field lines, allowing, in principle, high asymptotic Lorentz factors
of the polar outflow.
Pressure support by the external medium  may give rise to collimation of the central outflow (Bromberg \& Levinson 2007; Lyubarsky 2009).
Magnetic outflows have been considered also in the context of the unexpected paucity of optical flashes seen in 
GRBs (e.g., Zhang \& Kobayashi 2005; Mimica et al. 2009), and it has been shown that even moderate magnetization ($\sigma\sim0.3$) 
would strongly suppress the reverse shock.   The question remains as to how the internal shocks that produce the prompt emission form
under these conditions.   Alternative  explanations for the paucity of optical flashes, e.g., an early onset of a R-T instability (Levinson, 2009, 2010a), have been offered recently.

An key issue in the theory of magnetically dominated outflows is magnetic energy dissipation. 
The process by which magnetic energy is converted into kinetic energy has not been identified yet.
Stationary magnetic outflows allow, in general, only partial conversion of magnetic-to-kinetic energy.  For a split monopole field acceleration
ceases at an asymptotic Lorentz factor $\Gamma_\infty\sim\sigma_0^{1/3}$, and magnetization $\sigma_\infty\sim\sigma_0^{2/3}$ , 
where $\sigma_0$ is the initial magnetization of the expanding shell.   A better conversion
can be achieved if the outflow is collimated into a small opening angle, $\theta\simeq\Gamma^{-1}$ (Komissarov et al. 2009).  
Such small opening angles are problematic, particularly for the GeV bursts, for which $\Gamma_\infty\sim500-1000$ is inferred.
Corking from a star that, within the framework of the collapsar model, may be relevant to long bursts can alleviate the latter 
condition (Tchekhovskoy et al., 2010; Komissarov et al. 2010).  However, even then $\sigma_\infty$ cannot be much smaller 
than unity (Lyubarski 2010).  Furthermore, there is evidence for extreme Lorentz factors also in short bursts, where strong deconfinement 
is not anticipated. 

Recently it has been shown (Granot et al., 2010; hereafter GKS10; Lyutikov 2010a,b) that time-dependent effects 
may play a crucial role in the acceleration of a magnetized flow.   
Unlike a stationary flow, for which acceleration ceases at $\Gamma_\infty\sim\sigma_0^{1/3}$, an 
impulsive spherical shell expelled by a central source continues accelerating, even after loosing causal contact 
with the engine, until reaching nearly complete conversion of magnetic energy into bulk kinetic energy. 
The terminal Lorentz factor of a shell expanding in vacuum is $\Gamma_\infty\simeq \sigma_0$.  During the acceleration phase the major fraction of the 
shell energy is contained in a layer of width $2r_0$, where $r_0$ is the initial shell's width,
bounded between the front of a rarefaction wave reflected from the central source and the 
head of the shell.   The average Lorentz factor of the shell, roughly equals the Lorentz factor of the fluid at the rarefaction front, evolves as 
$<\Gamma>\propto t^{1/3}$.   Once the shell enters the coasting phase 
its width starts growing and its magnetization continues to drop, until nearly full conversion is accomplished.

However, for sufficiently high values of the initial magnetization $\sigma_0$ the evolution of the system is  significantly 
altered by the ambient medium  well before the shell reaches its coasting phase (Levinson 2010b).  The maximum Lorentz factor of
the shell is limited to values well below  $\sigma_0$. To be concrete, for a shell of initial energy $E=10^{52}E_{52}$ erg
and size $r_0=10^{12}T_{30}$ cm expelled into a medium having a uniform density $n_i$ the Lorentz factor is limited to
$\Gamma_{\rm max}\simeq180(E_{52}/T_{30}^3 n_i)^{1/8}$ in the high sigma limit.   The reverse shock and any internal shocks
that might form if the source is fluctuating are shown to be very weak.   The restriction on the Lorentz factor is more severe 
for shells propagating in a stellar wind.    Intermittent ejection of multiple, thin shells does not seem to help, as even in vacuum 
unlikely small duty cycle is required in order for shells to collide after reaching the coasting phase.
Such episodes are expected to produce a smooth, relatively fast rising slowly decaying (power law) light curve, even in a multi-shell scenario.
Events like GRB080916C and GRB090510 (Abdo et al. 2009a,b) are not easily accounted for by the impulsive high-sigma shell model.

The main conclusion from those recent studies of magnetic outflows is that processes beyond ideal MHD, as
might occur in e.g., a striped wind model,  are required.

\begin{figure}
\centering
\includegraphics[width=100mm,height=80mm]{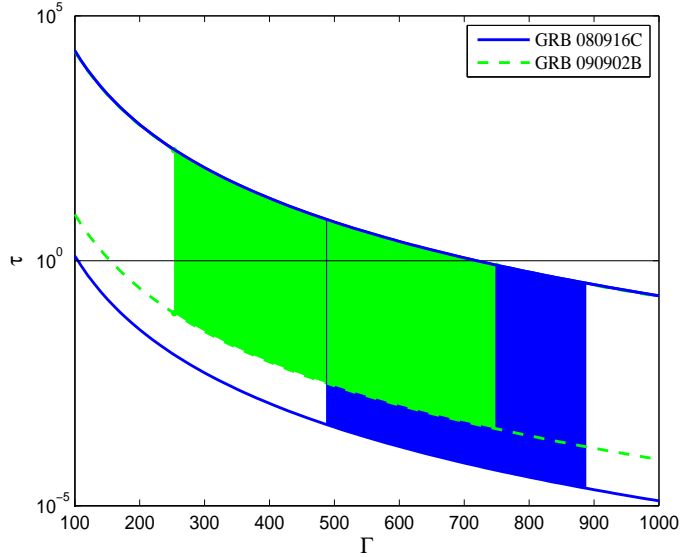}
\caption{\label{fig:a}Optical depth at the shells' collsion radius as a function of Lorentz factor, for different values of
$\delta t$ - the time interval between ejections of the two colliding shells.  The upper curves correspond to the dynamical time 
of a 3 solar mass black hole, $\delta t\simeq10^{-4}$ s.  The lower curves are limits from observations, with $\delta t=0.5$ s for
GRB080916C, and $\delta t=0.05$ s for GRB090902B.   The range of $\Gamma$ in each source represents various estimates 
taken from the literature (Bromberg et al. 2010).}
\end{figure}
\subsection{Thermal emission and relativistic radiation mediated shocks}
Shocks that form by overtaking 
collisions can dissipate energy at radii $r_d>\Gamma^2c\delta t$, where $\Gamma$ is the Lorentz factor of the slow shell
and $\delta t\ge r_s/c$ is time interval between ejections of the two consecutive shells. In blazars and microquasars with $\Gamma\sim 1-50$ 
dissipation by internal shocks is expected close to the BH, consistent with (but not necessarily implied by) the 
short durations of strong flares observed in these objects, particularly in TeV blazars.  

In GRBs $r_d>10^5 r_s$ 
or so for the Lorentz factors envisaged.  A rough condition for internal shocks to form above the photosphere 
is $\Gamma>200 L_{52}^{1/5}\delta t^{-1/5}_{-3}$ , where $L_{52}$ is the burst luminosity in units of $10^{52}$ erg/s, and 
$\delta t_{-3}=\delta t/(1 {\rm ms})$.
Results of more exact calculations (Bromberg et al. 2010) are exhibited in figure \ref{fig:a}, where the Thompson optical
depth at the radius of shock formation, $\tau$, is plotted agains the flow Lorentz factor $\Gamma$.  The two colored 
areas correspond to a range of $\delta t$ and $\Gamma$ for 
GRB 080916C (blue) and GRB 090902B (green), as explained in the figure caption. 

As shown, in case of GRB 080916C internal shocks are expected to form above the photosphere for shells accelerated to $\Gamma>700$.
Such shocks are probably collisionless, and can Fermi accelerate particles to nonthermal energies.  
Slower shells ejected over time intervals of the order of the dynamical time will collide below the photosphere, at a moderate optical depth
$\tau<100$.   The parameter space explored in Figure \ref{fig:a} indicates that in this burst the majority of the explosion energy dissipates above the photosphere, behind collisionless shocks, so that 
a thermal component may be very weak or absent.  

In GRB 090902B, on the other hand, the major fraction of the energy is likely to dissipate below the photosphere, albeit in a region 
of moderate optical depth, $\tau<300$, as seen in Figure \ref{fig:a}.  Only sufficiently 
slow shells that have large separations, such  $r_s/\delta t$ is much larger than the anticipated duty cycle of the engine, will
collide above the photosphere to form collisionless shocks. 
The shocks that form below the photosphere, where the Thomson depth exceeds unity, 
are mediated by Compton scattering (Bromberg \& Levinson 2008; Katz et al. 2010). 
Under conditions anticipated in GRBs, such relativistic radiation mediated shocks (RRMS) 
convect enough radiation upstream to render photon production in the shock transition negligible (Bromberg, et al., 2010),
unlike the case of shock breakout in supernovae and hypernovae (Katz et al. 2010).  Bulk Comptonization then produces a relatively 
low thermal peak, followed by a broad, nonthermal component
in the immediate downstream that extends up to the KN limit in the shock frame 
(or up to  $\sim \Gamma m_e c^2$ in the observer frame), and perhaps even beyond (Budnik et al. 2010), 
depending on details.   The radiation produced downstream is trapped for times much longer than the shock crossing time,
and diffuses out after the shock breaks out of the photosphere and becomes collisionless.   At what depth downstream equilibrium is established 
is yet an open issue.  Since the enthalpy downstream is dominated by radiation, a full equilibrium  is not expected for the moderate 
optical depths found above.   We contend that a thermal component plus a hard tail, as seen in GRB 090902B, can be emitted from the
downstream of a RRMS. A full treatment requires detailed transfer calculations, or Monte-Carlo simulations.  

\acknowledgement

I thank Omer Bromberg and Yuri Lyubarsky for enlightening discussions.  This work was supported by 
an ISF grant for the Israeli Center for High Energy Astrophysics.

\end{document}